\documentclass{ecai}

\usepackage{graphicx}
\usepackage{latexsym}
\usepackage{times}
\usepackage{soul}
\usepackage{url}
\usepackage[hidelinks]{hyperref}
\usepackage[utf8]{inputenc}
\usepackage[small]{caption}
\usepackage{graphicx}
\usepackage{amsmath}
\usepackage{amsthm}
\usepackage{booktabs}
\usepackage{algorithm}
\usepackage{algorithmic}
\usepackage{amsfonts}
\usepackage{textcomp}
\usepackage[table,xcdraw]{xcolor}
\usepackage{multirow}
\usepackage{bm}
\usepackage{arydshln}
\usepackage{bbding}

\begin{document}

\begin{frontmatter}

\title{Antecedent Predictions Are More Important Than You Think: An Effective Method for Tree-Based Code Generation}


\author[A]{\fnms{Yihong}~\snm{Dong}}
\author[A]{\fnms{Ge}~\snm{Li}\thanks{Corresponding Author. Email: lige@pku.edu.cn}}
\author[A]{\fnms{Xue}~\snm{Jiang}}
\author[A]{\fnms{Zhi}~\snm{Jin}}

\address[A]{Key Lab of High Confidence Software Technology, MoE (Peking University)}

\begin{abstract}
Code generation focuses on automatically converting natural language (NL) utterances into code snippets. 
Sequence-to-tree (Seq2Tree) approaches are proposed for code generation with the aim of ensuring grammatical correctness of the generated code. These approaches generate subsequent Abstract Syntax Tree (AST) nodes based on the preceding predictions of AST nodes. However, existing Seq2Tree approaches tend to treat both antecedent predictions and subsequent predictions equally, which poses a challenge for models to produce accurate subsequent predictions if the antecedent predictions are incorrect under the constraints of the AST.
Given this challenge, it is necessary to pay more attention to antecedent predictions compared to subsequent predictions. To this end, this paper proposes a novel and effective method, named Antecedent Prioritized (AP) Loss, which prioritizes antecedent predictions by leveraging the position information of the generated AST nodes.
We design an AST-to-Vector (AST2Vec) method that maps AST node positions to two-dimensional vectors, thereby modeling the position information of AST nodes. To evaluate the effectiveness of our proposed loss, we implement and train an Antecedent Prioritized Tree-based code generation model called APT. Experiments on four benchmark datasets demonstrate that with better antecedent predictions and accompanying subsequent predictions, APT achieves significant improvements, indicating the superiority and generality of our proposed method.
\end{abstract}

\end{frontmatter}

\section{Introduction}
Code generation is an essential generation task in the field of natural language processing (NLP) and software engineering, which deals with automatically generating a piece of executable code from NL utterances. In recent years, a series of Seq2Tree models have made remarkable achievements for code generation \cite{DongL16,LapataD18,RabinovichSK17,Yin17,TranX,Yin19Reranking,Shin19,SunZMXLZ19,TreeGen,Xie21MLTranX,Jiang21TranXRL,JiangSGMYS22}. 
Specifically, given an NL utterance input, instead of outputting a sequence of code tokens directly, the Seq2Tree model outputs a sequence of AST actions. These AST actions correspond to AST nodes in a traversal method based on the abstract syntax description language (ASDL) grammar, and can eventually be converted into code via a certain deterministic function \footnote{According to the compiler theory, AST nodes can be transformed into code by extracting all leaf nodes and some key nodes of AST, which is the deterministic function.}. Depending on AST structure and ASDL grammar, Seq2Tree models ensure the grammatical correctness of generated code. Furthermore, AST structure and ASDL grammar also help Seq2Tree models shrink the search space, i.e., antecedent predictions will constrain subsequent predictions, to generate well-formed codes.

Despite the success of the above models, they still neglect that antecedent predictions play an essential role in Tree-based code generation. The AST corresponding to code is a structure of progressive derivation from the root node to leaf nodes, and child nodes are derivations of parent nodes. Under the constraints of AST structure and ASDL grammar, parent nodes can easily determine the type, order, and even number of child nodes. When the output node sequence of Seq2Tree models is based on the pre-order traversal of AST\footnote{Note that breadth-first traversal has the same conclusion, and we use pre-order traversal as examples in this paper.}, the parent node is always ahead of its child nodes. Consequently, when the antecedent parent node is wrongly generated, Seq2Tree models can barely predict subsequent child nodes correctly, which dramatically impacts the effectiveness of Seq2Tree models. 

\begin{figure}[t!]
\centering
\includegraphics[width=0.48\textwidth]{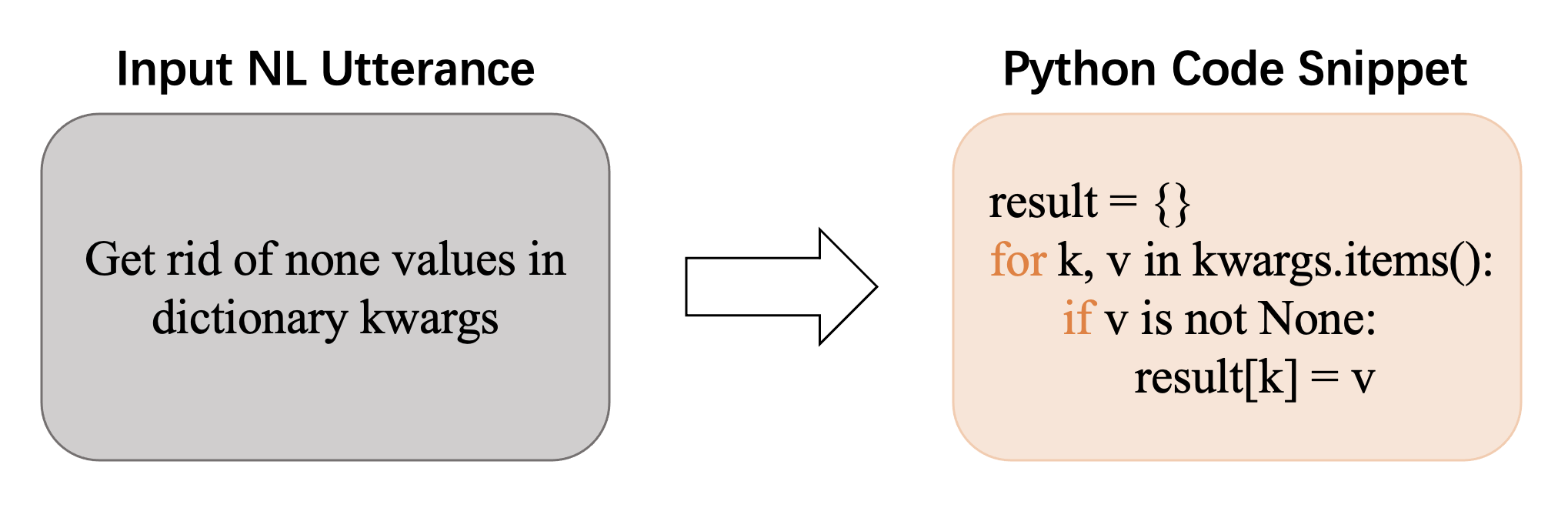} 
\caption{
An illustration of code generation: Given a NL utterance as input, the task is to convert it into corresponding code snippet.
}
\label{example}
\end{figure}

In this paper, we propose a novel loss function, namely AP Loss, that helps models prioritize antecedent predictions during training, so that generating more accurate antecedent predictions and accompanying subsequent predictions during inference. Intrinsically, AP Loss is a dynamically scaled cross entropy loss containing scaling and magnitude factors. The scaling factor is related to the position of the traversal order of AST nodes, and the magnitude factor controls the magnitude of the total loss function. Intuitively, AP Loss automatically reduces the contribution of susceptible subsequent predictions and focuses on more critical antecedent predictions during training. Therefore, we propose an AST2Vec method that models AST nodes with vectors to distinguish between antecedent nodes and subsequent nodes in AST. We implement and train an Antecedent Prioritized Tree-based code generation model called APT, utilizing the AP Loss. To demonstrate the efficacy of our proposed method, we conduct extensive experiments on several benchmark datasets. Experimental results show that APT outperforms the state-of-the-art (SOTA) Seq2Tree methods, including ML-TRANX \cite{Xie21MLTranX}, TRANX-RL \cite{Jiang21TranXRL}, and ASED \cite{JiangSGMYS22}, on four benchmark datasets. Furthermore, we note that AP Loss is suitable for not only different traversal methods but also different forms of weights, and maintains an advantage over Seq2Tree models, which indicates its generality.

The main contribution of this paper can be summarized as follows:
\begin{itemize}
    \item We propose that antecedent predictions are more critical than subsequent predictions in the tree-based code generation framework.
    
    \item We propose AP Loss, an effective method for Seq2Tree models, which prioritizes antecedent predictions by leveraging the position information of generated AST nodes. Additionally, we introduce the AST2Vec method, which represents the position information of AST nodes through two-dimensional vectors and has the capability to reconstruct the corresponding AST.
    
    \item We instantiate an Antecedent Prioritized Tree-based code generation model, referred to as APT. APT outperforms the SOTA Seq2Tree methods on four benchmark datasets. Extensive experimental results demonstrate the effectiveness and generality of our proposed method.
    
    
    
\end{itemize}

\section{Motivation}


\begin{figure}[t!]
\centering
\includegraphics[width=0.5\textwidth]{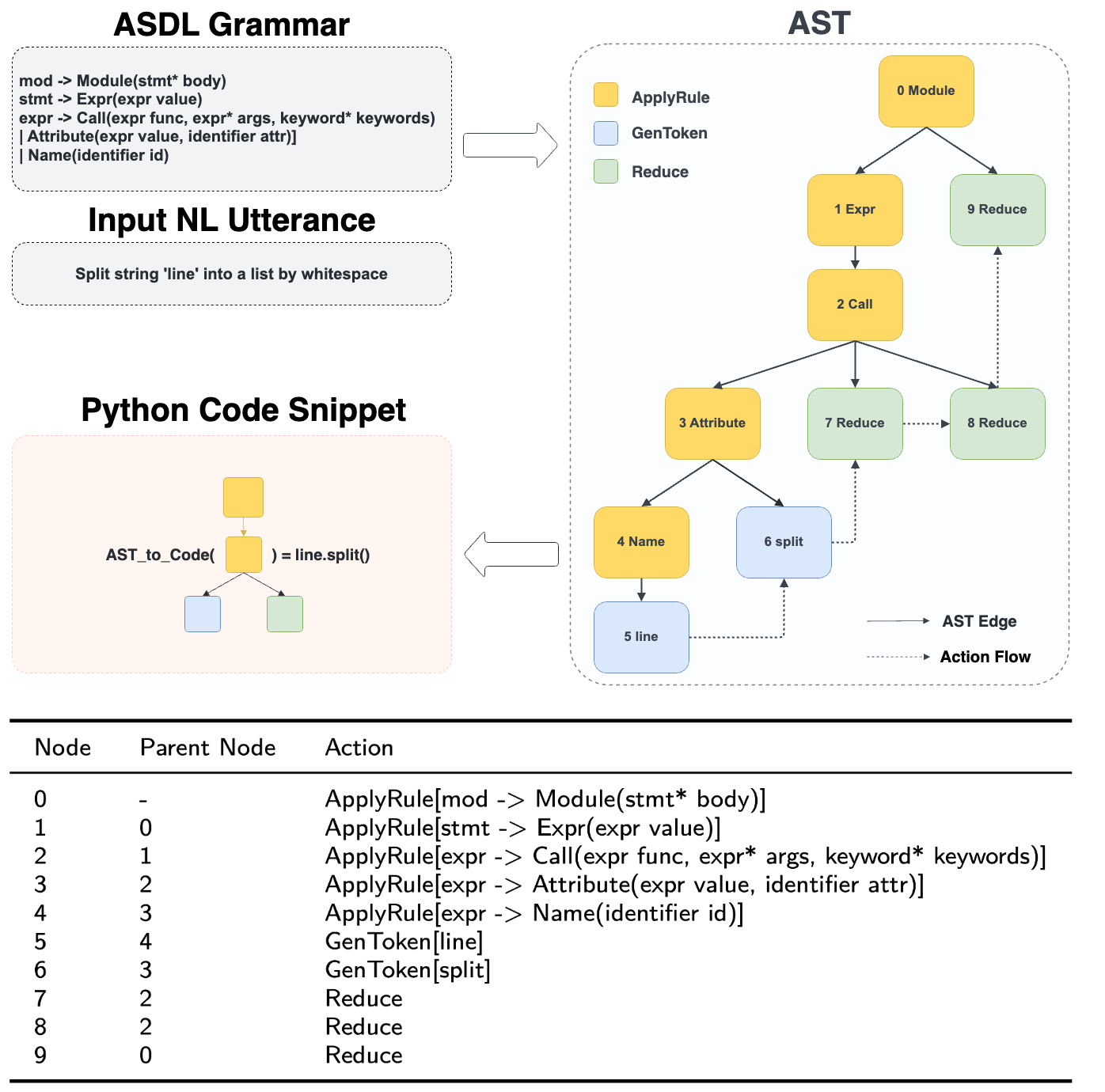}
\caption{
The procedure of code generation by Seq2Tree models. The order of nodes is based on the pre-order traversal of AST.  
}
\label{ast}
\end{figure}
Source code is specialized text with strict syntactic rules and specific structural patterns. Certain key statements significantly impact the structure of code. Some famous examples such as \textit{for}, \textit{while}, and \textit{if}, are shown in an example of code generation in Fig. \ref{example}.
When the code is converted to AST with ASDL grammar, these key statements are presented as dominant antecedent nodes \footnote{An antecedent node of a node is either the parent of this node or the parent of some ancestors of this node.} in the AST. 
An AST and its corresponding ASDL grammar are depicted in Fig. \ref{ast}. In this illustration, the yellow nodes represent antecedent nodes that determine the tree's structure. From this observation, we further summarize several properties that AST nodes exhibit under the constraints of ASDL grammar:
\begin{enumerate}
    \item The parent node can determine the left side of the production rule of the child node, e.g., since the right side of the production rule of node 0 is \textit{Module(stmt* body)}, the left side of the production rule of node 1 should be \textit{stmt}. 
    \item The parent node is able to control the number and order of the child node. For instance, the right side of the production rule of node 3 is \textit{Attribute(expr value, identifier attr)}, therefore node 3 has two child nodes, the first is \textit{expr} and the second is \textit{identifier}.
    \item The parent node allows for a decision of the type of the child node. For example, because the right side of the production rule of node 4 is \textit{Name(identifier id)}, the type of node 5 has to be the id of the identifier. Similarly, the type of node 6 has to be the attribute of the identifier.
\end{enumerate}
The properties of an AST underscore the critical role played by the antecedent nodes in code generation. Should the antecedent node's prediction prove inaccurate, subsequent node predictions will be adversely impacted, resulting in incorrect predictions.
Fig. \ref{ast} shows the procedure of converting an input NL utterance into the corresponding python code snippet by Seq2Tree models by generating actions to construct an intermediate representation AST.
During training, the Seq2Tree model generates the subsequent action predicated on the previous ground-truth action at each training step. Conversely, during testing, the model lacks knowledge of the ground truth action and proceeds to generate the next action based on the previously generated action. This discrepancy results in a non-negligible inconsistency between the training and inference phases. 
It is incredibly challenging to generate an accurate subsequent prediction based on incorrect antecedent predictions under AST structure and ASDL grammar constraints. Thus, the cost of producing wrong antecedent predictions is quite expensive during inference. To this end, in this paper, we propose an effective method that pays more attention to pre-generation during training and alleviates the negative effect of training inconsistency for Tree-based code generation methods.

\section{APT}
In this section, we first propose a method of representing AST node positions as vectors, called AST2Vec. Then, we describe the principle and architecture of our base model. Finally, we introduce AP Loss in detail. 

\begin{figure}[t!]
\centering
\includegraphics[width=0.5\textwidth]{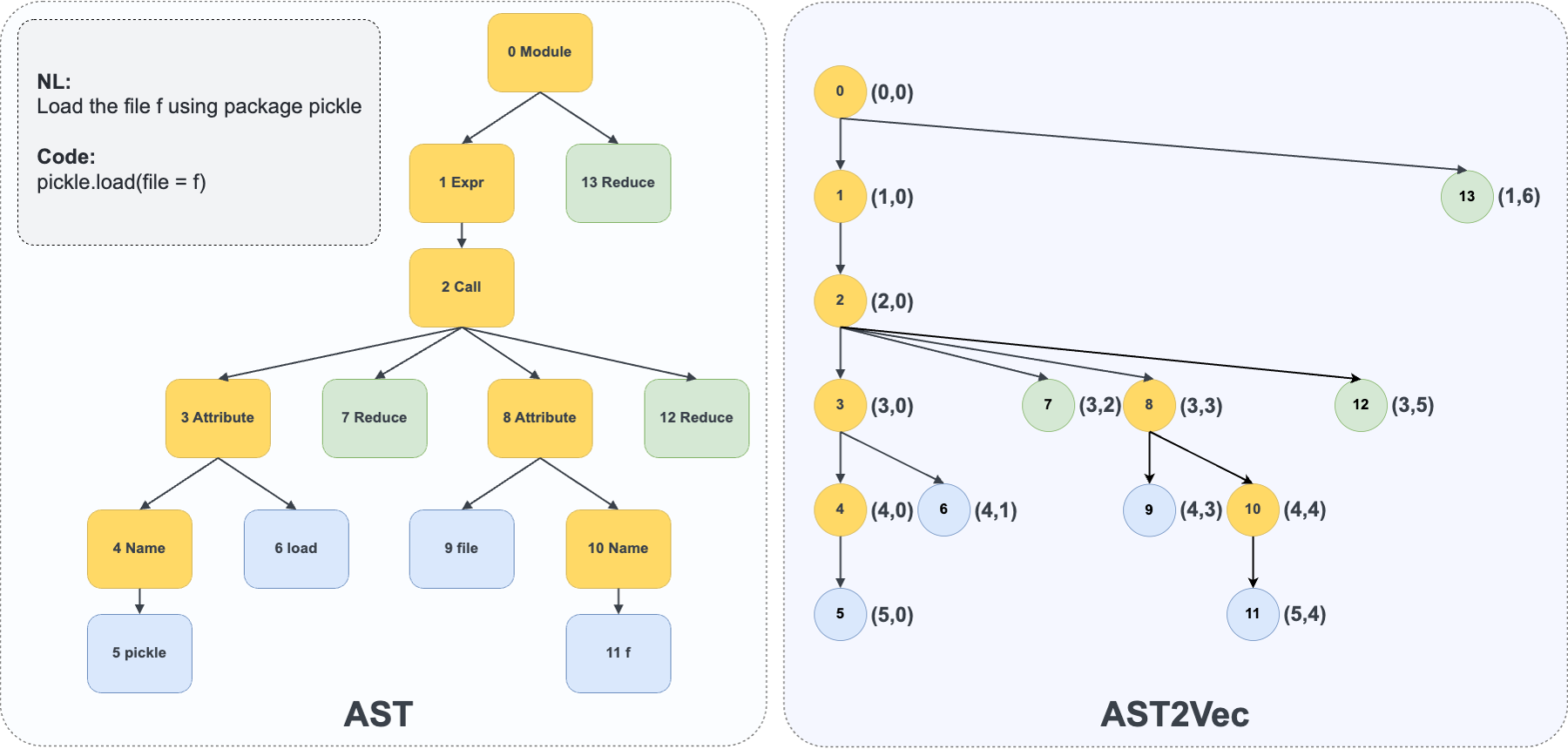}
\caption{
An Example of AST2Vec.
}
\label{ast2vec}
\end{figure}

\subsection{AST2Vec}
The sequence with pre-order traversal cannot represent the position information of AST nodes in AST well. Therefore, we consider adopting two-dimensional vectors to represent the position information of AST nodes. These vectors need to have the ability to reconstruct AST, which has to comply with the following rules:
\begin{itemize}
    \item The vector corresponding to each AST node should be unique;
    \item The capability to determine a unique parent node for any child node based on vectors.
\end{itemize}
Therefore, we design an AST2Vec method that conforms to these two rules. Specifically, we perform a pre-order traversal for an AST starting from the root node, and the vertical coordinate of the child node is that of its parent node + 1. If the parent node of the current node is traversed for the first time, and its horizontal coordinate is the horizontal coordinate of its parent node. If the parent node of the current node is traversed for the second and more times, its horizontal coordinate is the maximum of horizontal coordinates of all traversed nodes + 1. We provide the pseudocode for our AST2Vec method in Algorithm \ref{algorithm1} and an example of AST2Vec is shown in Fig. \ref{ast2vec}.


Representing ASTs with vectors based on AST2Vec offers two significant advantages:
\begin{enumerate}
    \item \emph{We can reconstruct the AST according to the corresponding vectors.} According to the algorithm used to generate the following-sibling nodes, each node is assigned a unique vector that does not overlap with any other nodes. This is achieved by setting the horizontal coordinate of a node to the maximum horizontal coordinate of all preceding nodes plus one, while ensuring that child nodes of following-sibling nodes have greater vertical coordinates. Furthermore, to determine the parent node of a given node, we look for the node whose vertical coordinate is one less than that of the current node, and whose horizontal coordinate is the first one less than or equal to the current node's horizontal coordinate. For example, in Fig. \ref{ast2vec}, the parent node of nodes 9 and 10 must be node 8, rather than node 7 or node 3.
    \item \emph{The subtraction of vectors satisfies the Triangle Law \footnote{To find the difference of two vectors $\bm{b}_1-\bm{b}_2$ that are coinitial, just draw a vector from the tail of $\bm{b}_2$ to the tail of $\bm{b}_1$}, which can reflect the relationship between AST nodes.} Concretely, the vector connecting two nodes can be represented by calculating the difference between their respective vector representations. For instance, in Fig. \ref{ast2vec}, the vector representation for node 8 is (3, 3), while that for node 9 is (4, 3). Therefore, the vector from node 8 to node 9 is computed as follows: $(4, 3) - (3, 3) = (1, 0)$. 
\end{enumerate}

In conclusion, the unique vector representation assigned to each node ensures that the AST can be accurately reconstructed from the corresponding vectors, while also allowing for easy determination of parent-child relationships between nodes. Further, the use of vector subtraction to calculate the difference between vectors satisfies the Triangle Law, allowing for an accurate reflection of the relationships between AST nodes. These benefits demonstrate the effectiveness of the AST2Vec approach in capturing the structural information of ASTs in a concise and useful way.

\begin{algorithm}[t!]
\caption{AST2Vec}\label{algorithm1}
\begin{algorithmic}[1]

\REQUIRE{The AST $\mathcal{T}$ or the pre-order traversal action sequence of AST $\mathbf{a}$.}
\ENSURE{The vectors $\bm{v}$.}

\STATE Initial the vectors $\bm{v}$, the record set $R = \emptyset$, and the maximum $maxn = 0$.
\IF{$\mathbf{a}$ is not exist} 
\STATE $\mathbf{a} \leftarrow $ the pre-order traversal of $\mathcal{T}$.
\ENDIF
\FOR{$a_t$ in $\mathbf{a}$}
\IF{$a_t$ is root node} 
\STATE $v_t = (0, 0)$.
\STATE continue.
\ENDIF
\STATE $a_p \leftarrow$ the parent node of $a_t$. 
\IF{$a_p \notin R$} 
\STATE $v_t = (v_p.d_1 + 1, v_p.d_2)$. 
\STATE add $a_p$ to $R$.
\ELSE
\STATE $maxn = maxn + 1$.
\STATE $v_t = (v_p.d_1 + 1, maxn)$.
\ENDIF
\ENDFOR
\RETURN{$\bm{v}$}
\end{algorithmic}
\end{algorithm}

\subsection{Base Model}
\label{Background}

Our proposed AP Loss is suitable for all Seq2Tree model. Here we adopt TRANX \cite{TranX} as our base model, given its widespread usage and demonstrated competitive performance in code generation.
 Numerous Seq2Tree methods are based on and improve upon TRANX, including ML-TRANX \cite{Xie21MLTranX}, TRANX-RL \cite{Jiang21TranXRL}, ASED \cite{JiangSGMYS22}, etc.

In Fig. \ref{ast}, TRANX first outputs a sequence of actions to construct AST in pre-order traversal based on ASDL grammar and the input NL utterance. Then a given function AST\_to\_Code(*) is called to map the generated intermediate AST into code. Concretely, three types of ASDL grammar-based actions are available:

    \textsc{\textbf{ApplyRule}[\textit{c}]} actions apply a constructor \textit{c} to the composite field of a node with the same type as \textit{c}, generating the child node using the fields in \textit{c}.
    
    \textsc{\textbf{Reduce}} actions complete the generation of the child node for a field that has optional($?$) or multiple($\star$) cardinalities.
    
    \textsc{\textbf{GenToken}[\textit{v}]} actions expand the field of a node to generate a token \textit{v}.

TRANX uses an attentional encoder-decoder framework with augmented recurrent connections to interpret the topology of AST. The encoder of TRANX is a Bidirectional Long Short-term Memory (BiLSTM) network, and the decoder of TRANX is an LSTM network. Given an input NL utterance $\mathbf{x}$ of $n$ tokens, i.e., $\left\{x_{i}\right\}_{i=1}^{n}$, the BiLSTM encoder learns vectorial representations $\{\mathbf{z}\}_{i=1}^{n}$. Then, at each step $t$, the LSTM decoder generates hidden state $\mathbf{s}_{t}$ as:
\begin{equation}
    \mathbf{s}_t=f_{\operatorname{LSTM}}([\mathbf{e}_{t-1} : \tilde{\mathbf{s}}_{t-1} : \mathbf{p}_t], \mathbf{s}_{t-1}),
\end{equation}
where $f_{\operatorname{LSTM}}$ is the LSTM decoder transition function, $[:]$ denotes the concatenation of vectors, $\mathbf{e}_{t-1}$ and $\mathbf{p}_t$ denote the embedding of previous action and parent action, respectively. We define the attentional vector $\tilde{\mathbf{s}}_{t}$ as:
\begin{equation}
    \tilde{\mathbf{s}}_{t} = \operatorname{tanh}(\textbf{W}_c[\mathbf{c}_t : \mathbf{s}_t]),
\end{equation}
where $\operatorname{tanh}$ is the hyperbolic tangent function, $\textbf{W}_c$ is a parameter matrix, $\mathbf{c}_t$ denotes the context vector weighted by encoding representations $\{\mathbf{z}\}_{i=1}^{n}$ using attention.

Finally, the probability of an \textsc{ApplyRule[\textit{c}]} action with embedding $\mathbf{e}_{c}$ at each step $t$ is:
\begin{equation}
p(a_t=\textsc{ApplyRule}[\textit{c}]  \mid  a_{<t}, \mathbf{x}) = \operatorname{softmax}(\mathbf{e}_{c}^T\textbf{W}\tilde{\mathbf{s}}_{t}),
\end{equation}
where $\operatorname{softmax}$ is the softmax function, $\mathbf{e}_{c}^T$ denotes the transpose of embedding $\mathbf{e}_{c}$, and $\textbf{W}$ is another parameter matrix. Note that \textsc{Reduce} is considered as a special \textsc{ApplyRule} action. For the \textsc{GenToken} action, its probability is defined as:
\begin{align*}
& p(a_t=\textsc{GenToken}[v]  \mid  a_{<t},\mathbf{x}) \\ & =   p(\operatorname{gen}  \mid  a_{t},\mathbf{x})p(v \mid \operatorname{gen},a_{t},\mathbf{x}) \\
& + p(\operatorname{copy} \mid a_{t},\mathbf{x})p(v \mid \operatorname{copy},a_{t},\mathbf{x}),
\end{align*}
where 
\begin{align*}
& p(\operatorname{gen}  \mid  a_{t},\mathbf{x}) = \operatorname{softmax}(\textbf{W}\tilde{\mathbf{s}}_{t}), \\
& p(\operatorname{copy} \mid a_{t},\mathbf{x}) = 1 - p(\operatorname{gen}  \mid  a_{t},\mathbf{x}),\\
& p(v \mid \operatorname{gen},a_{t},\mathbf{x}) = \operatorname{softmax}(\mathbf{e}_{v}^T\textbf{W}\tilde{\mathbf{s}}_{t}), \\
& p(x_i \mid \operatorname{copy},a_{t},\mathbf{x})=\operatorname{softmax}(\mathbf{z}_{i}^T\textbf{W}\tilde{\mathbf{s}}_{t}).
\end{align*}

Similar to other Seq2Tree models, given a corpus $(\mathbf{x}, \mathbf{a})$, TRANX is trained to minimize the cross entropy loss:
\begin{equation}
\mathcal{L}_{\operatorname{ce}}(\mathbf{x}, \mathbf{a}; \mathbf{\theta})=-\sum_{t=1}^{T} \log p\left(a_{t} \mid a_{<t}, \mathbf{x}; \mathbf{\theta}\right),
\label{ce}
\end{equation}
where $T$ is the total number of actions and $\mathbf{\theta}$ denotes parameters of the model. 

\subsection{Antecedent Prioritized Loss}
\label{Antecedent Prioritized Loss}

We extend the standard cross entropy loss \eqref{ce} to AP Loss by exploiting the position information of AST traversal order. Specifically, a scaling factor $f(\mathbf{a}, t)^{-\gamma}$ is added with respect to AST action sequence and the action index $t$, and the AP Loss is defined as:
\begin{align}
\mathcal{L}_{\operatorname{ap}}(\mathbf{x}, \mathbf{a}; \mathbf{\theta}) &=-\sum_{t=1}^{T} f(\mathbf{a}, t)^{-\gamma} \log p\left(a_{t} \mid a_{<t}, \mathbf{x}; \mathbf{\theta}\right),  \label{pl}\\
f(\mathbf{a}, t) & =\operatorname{Norm}(\operatorname{AST2Vec}(\mathbf{a})_t)
\end{align}
where the tunable hyperparameter $\gamma \geq 0$, the function $\operatorname{AST2Vec}$ uses Algorithm \ref{algorithm1}, and $\operatorname{Norm}(\operatorname{AST2Vec}(\mathbf{a})_t)$ indicates the norm of the vector $\operatorname{AST2Vec}(\mathbf{a})_t$.  The scaling factor $f(\mathbf{a}, t)^{-\gamma}$ decreases as $f(\mathbf{a}, t)$ increases, which helps models down-weight actions with large $f(\mathbf{a}, t)$ and thus focus training on actions with small $f(\mathbf{a}, t)$, i.e., antecedent actions. The rationale for this is that, when the AST is mapped to the corresponding vectors using AST2Vec, the vector norm of the parent node is always smaller than that of the child node, as shown in Fig. \ref{ast2vec}. This property enables us to prioritize training on antecedent actions by reducing the contribution of descendant actions through the scaling factor. In summary, the hyperparameter $\gamma$ provides control over the weighting of different actions in our model, and the scaling factor helps us prioritize antecedent actions by reducing the contribution of descendant actions during training. 

\begin{figure}[t!]
\centering
\includegraphics[width=0.45\textwidth]{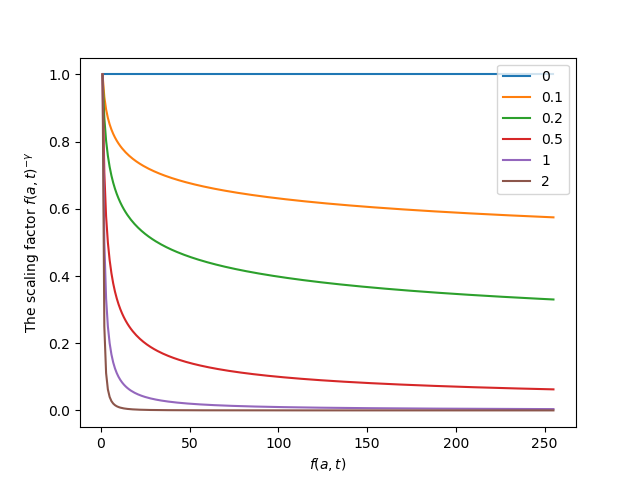} 
\caption{
The effect of varying $\gamma$ on the scaling factor $f(\mathbf{a}, t)^{-\gamma}$.
}
\label{r}
\end{figure}

\begin{figure}[h!]
\centering
\includegraphics[width=0.5\textwidth]{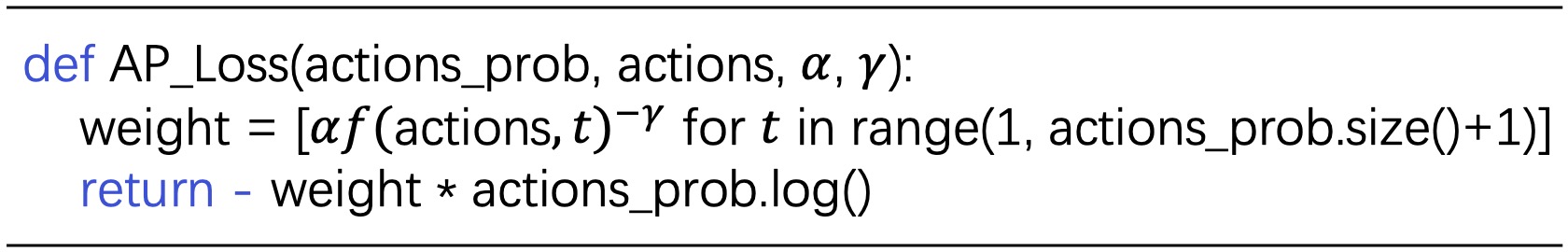} 
\caption{
Pseudocode for AP Loss.
}
\label{pc}
\end{figure}

Fig. \ref{r} displays the scaling factor for several values of $\gamma \in [0, 2]$. When $\gamma = 0$, the AP Loss ($\mathcal{L}_{\operatorname{ap}}$) reduces to the standard cross entropy loss ($\mathcal{L}_{\operatorname{ce}}$). It is advisable to avoid selecting $\gamma \textgreater 1$, as the resulting scaling factor $f(\mathbf{a}, t)^{-\gamma}$ can vary dramatically. For instance, when $\gamma = 2$, an action with index 10 would have a loss that is $100\times$ lower than that of $\mathcal{L}_{\operatorname{ce}}$. This could lead models to disregard subsequent predictions completely, which is undesirable. To achieve a balance between the loss of antecedent and subsequent predictions, a suitable scaling factor should be chosen. Thus, we recommend selecting $\gamma$ from the range 0.1 to 0.5. It is worth noting that implementing the AP Loss in code is relatively straightforward and we provide the pseudocode of AP Loss in Fig. \ref{pc}.

In practice, we insert an $\alpha$-adjusted magnitude factor in AP Loss:
\begin{equation}
\mathcal{L}_{\operatorname{ap}}(\mathbf{x}, \mathbf{a}; \mathbf{\theta})=-\alpha \sum_{t=1}^{T} t^{-\gamma} \log p\left(a_{t} \mid a_{<t}, \mathbf{x}; \mathbf{\theta}\right).
\end{equation}

We adopt the above form in our experiments because it yields improvements over the non-$\alpha$-adjusted form \eqref{pl}. Generally, $\alpha$ can be chosen from positive constants. 

\section{Experiment Setup}
\subsection{Datasets}

We conduct experiments on four public benchmark datasets as follows: \textbf{GEO} \cite{ZelleM96} consists of 880 US geographical NL utterances and their corresponding code defined in lambda logical forms. 
\textbf{ATIS} \cite{atis} is a collection of 5,410 paired flight data, which contains the NL description of flight information and their corresponding lambda calculus code like GEO.
\textbf{DJANGO} \cite{oda2015learning} is a popular code generation dataset consisting of NL-annotated python2 code extracted from Django Web framework.
\textbf{CONALA} \cite{YinDCVN08} contains 2,879 real-world data of manually annotated NL questions and their Python3 solutions on STACK OVERFLOW, which is more difficult because of its complex and extensive code composition. 

The detailed statistics of the above datasets are shown in Table \ref{statistics}.

\begin{table}[h!]
	\caption{Statistics of benchmark datasets. }	\label{statistics}
	\centering
        \scriptsize{
	\begin{tabular}{lllllll}
	    \toprule
        \multirow{2}{*}{Dataset}& \multicolumn{3}{c}{Examples Num} & \multicolumn{3}{c}{Avg Length} \\\cmidrule(r){2-4} \cmidrule(r){5-7}
        & Train & Dev & Test & NL & Code & AST \\
        \midrule
        GEO & 600 (480) & - (120) & 280 & 7.4 &  19.1 & 19.3 \\
        ATIS & 4,473 & 491 & 448 & 10.5 & 28.1 & 31.5\\
        DJANGO & 16000 & 1000 & 1805 & 14.1 & 10.6 & 14.4\\
        CONALA & 2,175 & 200 & 500 & 10.2 & 15.1 & 23.2\\
	    \bottomrule
	\end{tabular}}
  \end{table}

\subsection{Baselines}
To exhibit the effectiveness of our proposed method, we primarily compare APT with several competitive Seq2Tree methods, which can guarantee the syntactic correctness of generated code, as follows:

\textbf{SEQ2TREE} \cite{DongL16} uses a general attention-enhanced encoder-decoder model to generate logical forms by conditioning output trees.
\textbf{ASN} \cite{RabinovichSK17} outputs AST constructed by a decoder whose modular structure is dynamically determined, parallel to the structure of output trees.
\textbf{COARSE2FINE} \cite{LapataD18} first generates a rough meaning sketch abstracted from low-level information, and then fills in missing detail.
\textbf{TREEGEN} \cite{TreeGen} uses Transformer instead of LSTM to alleviate the long-dependency problem and adopts an AST reader to incorporate both grammar rules and AST structures.
\textbf{TRANX} \cite{TranX} is our base model which has detailed descriptions in Section \ref{Background}. TRANX and TRANX' are the versions of the pre-order and breadth-first traversals of TRANX, respectively.
\textbf{ML-TRANX} \cite{Xie21MLTranX} adopts a mutual learning framework to train models for different traversals based decodings jointly.
\textbf{TRANX-RL} \cite{Jiang21TranXRL} uses a context-based branch selector to dynamically determine optimal branch expansion orders for multi-branch nodes. 
\textbf{ASED} \cite{JiangSGMYS22} generates the current prediction with both the history and future information using an AST structure enhanced decoder. 

Following the previous work \cite{TreeGen,Xie21MLTranX,Jiang21TranXRL}, we mainly compare our proposed method to the SOTA Seq2Tree methods with similar parameter sizes to ensure the fairness of the comparison.

\subsection{Metrics}  
To evaluate the effectiveness of different methods, we use two widely-used evaluation metrics: the exact matching accuracy (EM) for all datasets and the corpus-level BLEU-4 for CONALA as a complement. The reason is that CONALA is a more challenging dataset compared to other datasets, EM may be too strict and can hardly reflect the performance of methods on it. 
As the AP Loss does not impact the inference time, the execution time of each method is not reported. It should be noted that none of the four publicly available benchmark datasets include test cases, thus the evaluation metric of "pass@k" is not applied.

\subsection{Settings}
We train our model for a maximum epoch of 80 with Adam \cite{adam} optimizer on a single GPU of Tesla V100-SXM2-32G. To ensure fair comparisons, we adopt the same experimental setup as TRANX \cite{TranX}. To be specific, we set the sizes of word embedding, action embedding, and hidden states as 128, 128, and 256, respectively. The beam size is set to 5 for GEO and ATIS, while 15 for DJANGO and CONALA. According to the analysis in Sections \ref{Antecedent Prioritized Loss} and \ref{selection}, we pick $\gamma \in [0.1, 0.5]$ using validation set and set $\alpha = 2$ for all datasets. To mitigate the instability of the model training, we exhibit the average performance of the model running five times.

\section{Experimental Results}




\begin{table}[h!]
	\caption{The performance of APT compared with various baselines.}	\label{table1}
	\centering
        \resizebox{0.49\textwidth}{!}{
	\begin{tabular}{llllll}
	    \toprule
        \multirow{2}{*}{Model}& GEO & ATIS & DJANGO & \multicolumn{2}{c}{CONALA}\\\cmidrule(r){2-2} \cmidrule(r){3-3}\cmidrule(r){4-4}\cmidrule(r){5-6}
         & EM & EM & EM & BLEU & EM\\
        \midrule
        SEQ2TREE  & 86.1 & 84.6 & - & - & -\\
        ASN  & 87.1 & 85.9 & - & - & -\\
        COARSE2FINE  & 88.2 & 87.7 & 74.1 & - & -\\
        TREEGEN  & 89.6 & 89.1 & - & - & - \\
        \hdashline
        \rowcolor[HTML]{C0C0C0} 
        TRANX  & $88.8 \pm 1.0$ &  $87.6 \pm 0.1$ & $77.3 \pm 0.4$ & $24.35 \pm 0.4$ & $2.5 \pm 0.7$\\
        ML-TRANX  & $89.2 \pm 0.6$ & $89.3 \pm 0.3$ & \bm{$79.6 \pm 0.3$} & $24.42 \pm 0.8$ & $2.2 \pm 0.4$\\
        TRANX-RL  & $89.5 \pm 1.2$ & $89.1 \pm 0.5$ & $77.9 \pm 0.5$ & $25.47 \pm 0.7$ & $2.6 \pm 0.4$\\
        ASED & $89.8 \pm 1.1$ & $88.9 \pm 0.7$ & $79.2 \pm 0.5$ & $25.56 \pm 0.6$ & $2.8 \pm 0.7$\\
        \midrule
        APT (ours) & \bm{$90.4 \pm 0.8$} & \bm{$89.8 \pm 0.7$} & $\bm{79.6 \pm 0.5}$& \bm{$28.02 \pm 0.7$} & \bm{$2.9 \pm 0.7$}\\
	\bottomrule
	\end{tabular}}
  \end{table}

\subsection{APT vs. Various Baselines}
Table \ref{table1} lists the comparison of APT and various baselines on four benchmark datasets.  
With the help of AP Loss, our APT achieves significant improvements compared with base modal TRANX \cite{TranX} across all datasets. In particular, APT relatively improves 15.1\% BLEU on CONALA dataset compared to TRANX, which demonstrates the gain effect of AP loss. APT surpasses all TRANX variants and other Seq2Tree models, demonstrating notable performance. It is important to highlight that EM is a stringent evaluation metric, and the improvements achieved by APT are more pronounce than previous works. As a result, these experimental results indicate that letting the Seq2Tree model concentrate on the predictions of the antecedent AST nodes during training helps code generation in inference. 

\begin{figure}[h!]
\centering
\includegraphics[width=0.48\textwidth]{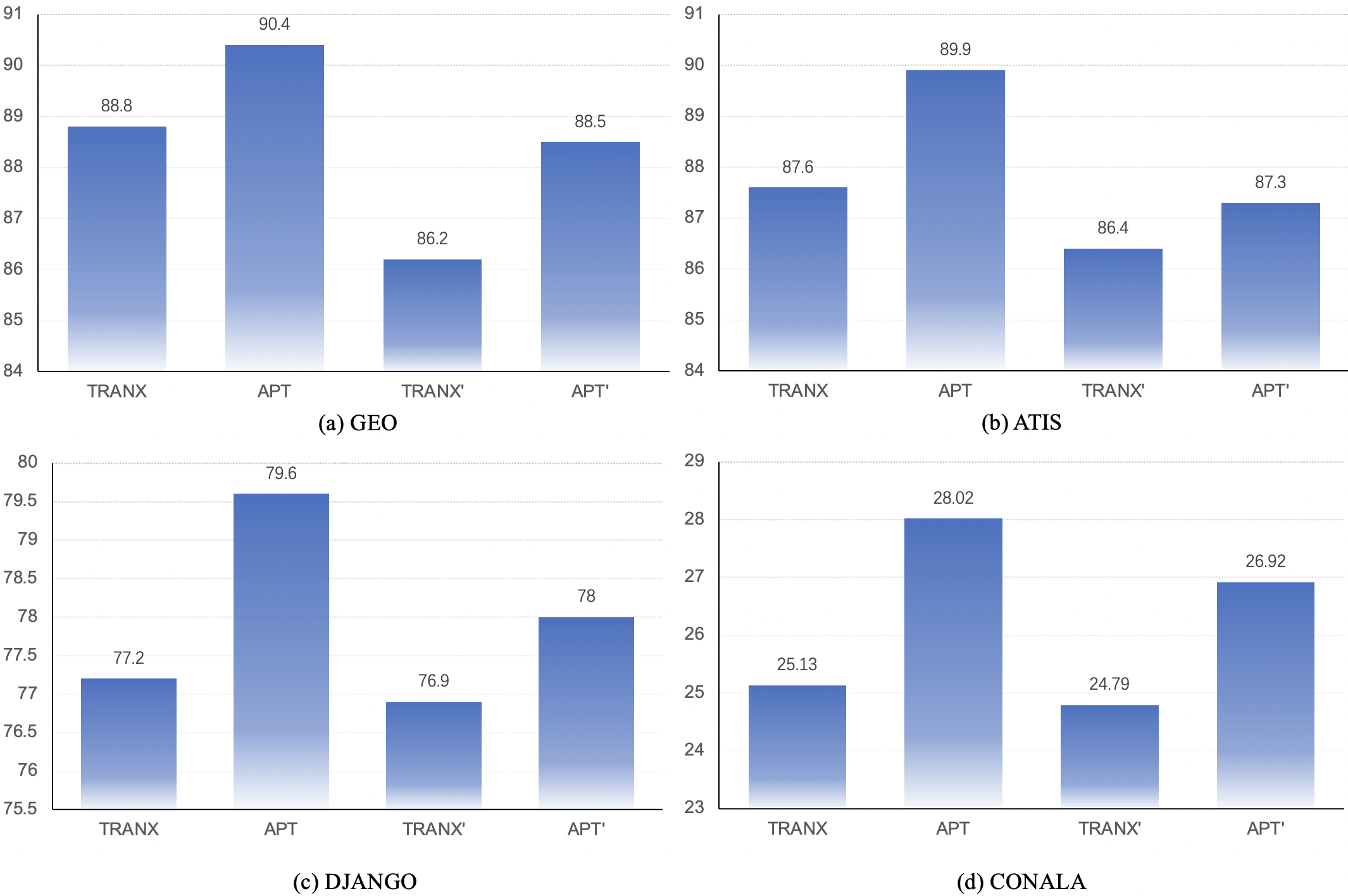}
\caption{
The performance of AP Loss based on the pre-order and breadth-first traversals of TRANX, i.e., APT and APT'. (BLEU for CONALA and EM for other datasets)
}
\label{PandB}
\end{figure}

\subsection{Pre-order vs. Breadth-first Traversal}
As shown in Fig. \ref{PandB}, the performance of APT' (breadth-first traversals of TRANX based on AP Loss) is inferior to that of APT on all datasets, due to the difference in the respective traversal method of the base model. A possible reason is that the breadth-first traversal has some drawbacks. For example, back to Fig. \ref{ast}, the attribute of the variable is output before the variable in breadth-first traversal, which increases the difficulty of prediction. Moreover, in Fig. \ref{ast}, the breadth-first traversal, compared to the pre-order traversal, prioritizes the output of the Reduce actions and delays the production of GenToken actions, which is actually evaluated by metrics such as BLEU and EM. The code token belongs to the leaf node of AST and is generated last in the breadth-first traversal, which dramatically affects the effect. Therefore, we should trade off both antecedent and leaf nodes in the generation. Nevertheless, both the pre-order and breadth-first traversals of TRANX are significantly improved using AP Loss. It implies that AP Loss can be applied to other reasonable traversal orders.

\begin{figure}[h!]
\centering
\includegraphics[width=0.48\textwidth]{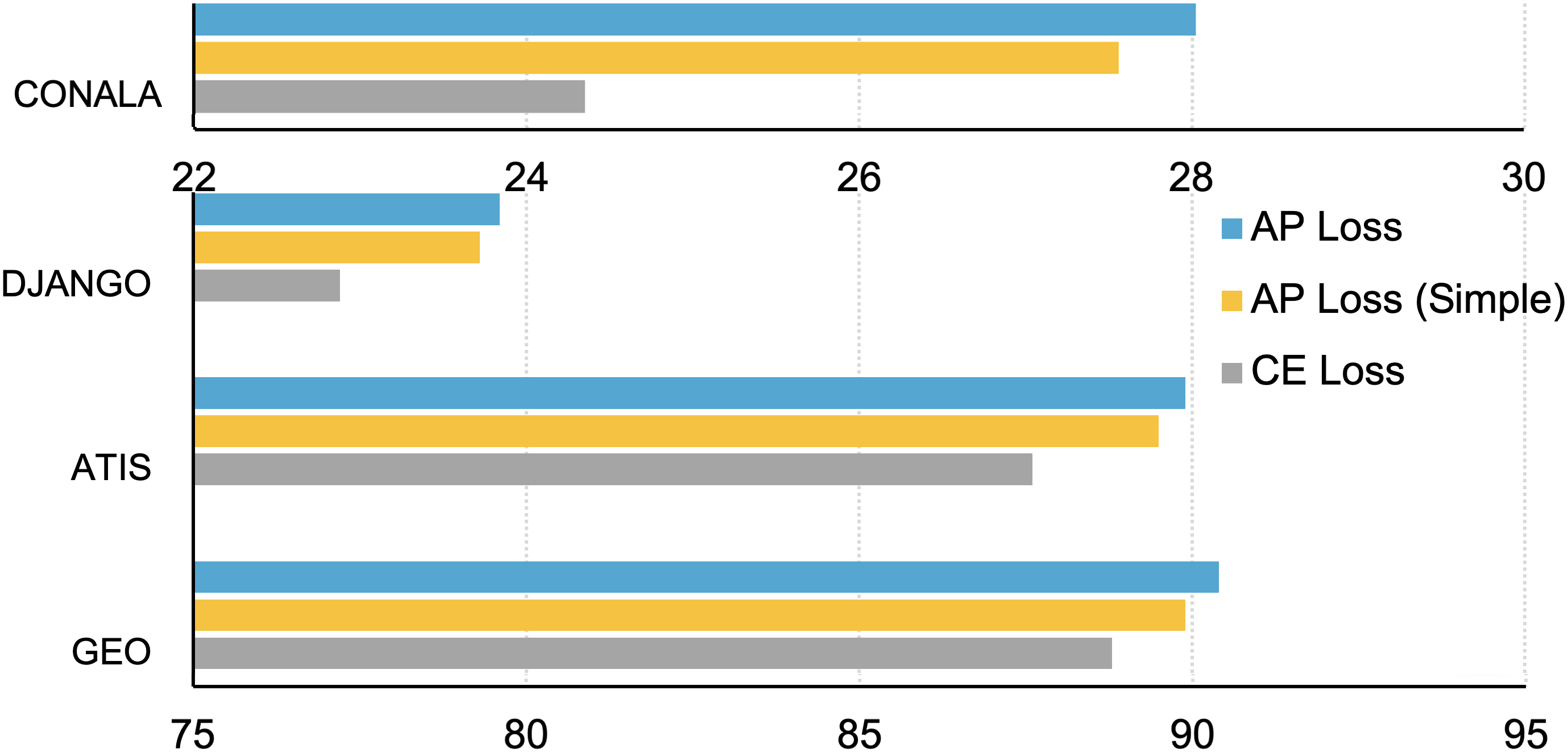}
\caption{
The performance of the base model based on AP Loss with different forms, where AP Loss (Simple) indicates $f(\mathbf{a}, t) = t$ in Eq. \eqref{pl}. (BLEU for CONALA and EM for other datasets)
}
\label{APL}
\end{figure}

\subsection{AP Loss vs. AP Loss (Simple)}
Note that $f(\mathbf{a}, t)$ in Eq. \eqref{pl} can be any reasonable function reflecting the effect of the antecedent prediction. For example, $f(\mathbf{a}, t) = t$ \footnote{In pre-order traversal of AST node, the parent node precedes the child node.}. We show the performance of the base model based on AP Loss with different forms for training in Fig. \ref{APL}. Remarkably, utilizing the simplified function $f(\mathbf{a}, t) = t$, which represents the AP Loss (Simple), still achieves a remarkable improvement compared to $\mathcal{L}_{\operatorname{ce}}$ on all datasets. This illustrates the importance of helping the model attach importance to antecedent predictions. We also find that the effect of using AP loss is significantly better than using AP Loss (Simple). This is because the position of the node in the pre-order traversal of AST only partially reflects its importance. The weight of sibling nodes in AP Loss (Simple) can vary widely, while that of sibling nodes in AP Loss is much smaller. 

\begin{table}[h!]
	\caption{Comparisons of Seq2Tree models and their use of AP Loss on CONALA.}	\label{percent2}
	\centering
        \small{
	\begin{tabular}{lll}
	    \toprule
        Model & BLEU & EM\\
        \midrule
        TRANX & $24.35 \pm 0.4$ & $2.5 \pm 0.7$\\
        + AP Loss & $28.02 \ (\uparrow 15.1\%) \pm 0.7$ & $2.9 \pm 0.7$\\
        \midrule
        ML-TRANX \cite{Xie21MLTranX} & $24.42 \pm 0.8$ & $2.2 \pm 0.4$\\
        + AP Loss & $27.96  \pm 0.7$ & $2.6 \pm 0.5$\\
        \midrule
        TRANX-RL \cite{Jiang21TranXRL} & $25.47 \pm 0.7$ & $2.6 \pm 0.4$\\
        + AP Loss & $27.88  \pm 0.6$ & $2.7 \pm 0.5$\\
        \midrule
        ASED \cite{JiangSGMYS22} & $25.56 \pm 0.6$ & $2.8 \pm 0.7$ \\
        + AP Loss & $27.91 \pm 0.7$ & $2.9 \pm 0.9$\\  
	    \bottomrule
	\end{tabular}}
 \end{table}

\subsection{AP Loss with different Seq2Tree methods}
We pick CONALA for the follow-up comparison of AP Loss with two reasons: 1) Python used by CONALA has a richer syntactic structure than Lambda code used by GEO and ATIS. 2) The average AST length of CONALA exceeds that of DJANGO by 60\%.

In Table \ref{percent2}, we apply AP Loss to Seq2Tree models and compare their performance on CONALA. The experimental results demonstrate that all these Seq2Tree models achieve significant improvements using AP Loss. Moreover, the effect of AP loss on four Seq2Tree models is quite similar, probably because 1) we uniformly adopt the hyperparameter $\gamma$ used for TRANX without adjusting; 2) improved model performance is usually accompanied by better antecedent predictions.

\begin{table}[h!]
	\caption{EM of the different percent of AST node sequences and code token sequences on CONALA. }	\label{percent}
	\centering
        \resizebox{0.49\textwidth}{!}{
	\begin{tabular}{lllllllll}
	    \toprule
        \multirow{2}{*}{Model}&\multicolumn{2}{c}{5\%}&\multicolumn{2}{c}{10\%} &\multicolumn{2}{c}{20\%}&\multicolumn{2}{c}{50\%}\\
        \cmidrule(r){2-3}\cmidrule(r){4-5}\cmidrule(r){6-7}\cmidrule(r){8-9}
        & AST & code & AST & code & AST & code & AST & code\\
        \midrule
        TRANX &	45.2 & 17.4 & 26.4 & 8.6 & 11.8 & 4.2 & 5.8 & 2.6 \\
        + AP Loss & 49.0 & 23.0 & 30.6 & 14.6 & 14.8 & 7.0 & 9.2 & 4.8 \\
        \midrule
        ML-TRANX  & 46.8 & 18.2 & 27.6 & 9.6 & 12.4 & 3.8 & 5.6 & 2.4 \\
        + AP Loss & 49.8 & 22.8 & 30.8 & 14.0 & 14.6 & 6.6 & 8.8 & 4.4 \\
        \midrule
        TRANX-RL  & 47.0 & 20.0 & 28.2 & 11.0 & 13.4 & 4.6 & 5.4 & 3.0 \\
        + AP Loss & 48.8 & 22.6 & 30.2 & 13.6 & 14.8 & 6.6 & 8.8 & 4.6 \\
        \midrule
        ASED  & 47.8 & 17.2 & 28.6 & 9.2  & 13.2 & 4.8 & 5.8 &  3.2 \\
        + AP Loss & 49.8 & 22.8 & 29.8 &  13.8 & 15.4 & 6.8  & 8.6 & 4.8 \\  
	    \bottomrule
	\end{tabular}}
 \end{table} 

In Table \ref{percent}, we test the EM of antecedent AST node and code token sequence predictions on CONALA. The experimental results demonstrate that Seq2Tree methods achieve higher EM on the first 5\%, 10\%, 20\%, and 50\% of AST node sequences and code token sequences using AP Loss. It indicates that the Seq2Tree method can obtain better antecedent predictions with the help of AP Loss. 

 \begin{table}[h!]
	\caption{Comparison of schedule sampling and AP Loss on CONALA, where p denotes the probability of using the label as input (TRANX is equivalent to the case of p=1), ED denotes exponential decay, e denotes the base of the exponent, and LD denotes linear decay.}	\label{ss} 
	\centering
        \scriptsize{
	\begin{tabular}{llcccc}
	    \toprule
        \multicolumn{2}{c}{\multirow{2}{*}{Model}}&\multicolumn{2}{c}{Test Set}&\multicolumn{2}{c}{Training Set}\\\cmidrule(r){3-4}\cmidrule(r){5-6}
        \multicolumn{2}{c}{} & BLEU & EM & BLEU & EM\\
        \midrule 
        \multicolumn{2}{c}{TRANX} & 24.35 & 2.5 & 63.09 & 36.11\\
        \midrule 
        \multicolumn{1}{c}{\multirow{4}{*}{+ schedule sampling}} & p=0 & 22.94 & 1.6 & 61.44	& 37.90\\
        & ED (e=0.9) & 24.45 & 1.6 & 65.22 & 36.41\\
        & ED (e=0.99) & 25.59 & 2.2 & \textbf{76.59} & \textbf{53.63}\\
        & LD & 25.14 & 2.0 & 66.85 & 37.95\\
        \midrule 
        \multicolumn{2}{c}{+ AP Loss} & \textbf{28.02} & \textbf{2.9} & 68.52 & 44.35\\
	    \bottomrule
	\end{tabular}}
 \end{table}

\begin{table*}[t!]
\caption{Case study examples on four benchmark datasets. The incorrect codes are marked in blue, while the first token of the counterparts is marked in red.}	\label{table3}
\centering
\footnotesize{
\begin{tabular}{llll}
    \toprule
    Dataset & Model & NL & Code\\
    \midrule
    \multirow{8}{*}{GEO} & \multirow{2}{*}{TRANX} & \multirow{4}{2.5cm}{What length is the r0?} & \multirow{2}{11.5cm}{\textcolor{blue}{r0} \textcolor{blue}{\XSolidBrush}}\\ \\
    & \multirow{2}{*}{APT} &  & \multirow{2}{11.5cm}{\textcolor{red}{(} len:i r0 ) \textcolor{red}{\CheckmarkBold}}\\ \\
    \cdashline{2-4}
    & \multirow{2}{*}{TRANX} & \multirow{4}{2.5cm}{What is the highest elev of c0?} & \multirow{2}{11.5cm}{( argmax \$0 ( and ( place:t \$0 ) ( loc:t \$0 c0 ) ) \textcolor{blue}{c0} ) \textcolor{blue}{\XSolidBrush}}\\ \\
    & \multirow{2}{*}{APT} &  & \multirow{2}{11.5cm}{( argmax \$0 ( and ( place:t \$0 ) ( loc:t \$0 c0 ) ) \textcolor{red}{(} elevation:i \$0 ) ) \textcolor{red}{\CheckmarkBold}}\\ \\
    \midrule
    \multirow{8}{*}{ATIS} & \multirow{2}{*}{TRANX} & \multirow{4}{2.5cm}{Is there ground transport from the ap0 into ci0 town?} & \multirow{2}{11.5cm}{( lambda \$0 e ( \textcolor{blue}{exists \$1 ( and ( = ( arrival\_time \$1 ) \$0 ) ( approx\_arrival\_time \$1 ti0 ) ( ground\_transport \$1 ) ( to\_city \$1 ci0 ) )} ) ) \textcolor{blue}{\XSolidBrush}}\\ \\
    & \multirow{2}{*}{APT} &  & \multirow{2}{11.5cm}{( lambda \$0 e ( \textcolor{red}{and} ( from\_airport \$0 ap0 ) ( ground\_transport \$0 ) ( to\_city \$0 ci0 ) ) ) \textcolor{red}{\CheckmarkBold}}\\ \\
    \cdashline{2-4}
    & \multirow{2}{*}{TRANX} & \multirow{4}{2.5cm}{Tell me about the type of aircraft call an ac0.} & \multirow{2}{11.5cm}{\textcolor{blue}{( lambda \$0 e ( exists \$1 ( and ( flight \$1 ) ( = ( aircraft\_code \$1 ) \$0 ) ( aircraft\_code \$1 ac0 ) ) ) ) \XSolidBrush}}\\ \\
    & \multirow{2}{*}{APT} &  & \multirow{2}{11.5cm}{\textcolor{red}{ac0 \CheckmarkBold}}\\ \\
    \midrule
    \multirow{8}{*}{DJANGO} & \multirow{2}{*}{TRANX} & \multirow{4}{2.5cm}{Is length of list\_ equals integer 0?} & \multirow{2}{11.5cm}{\textcolor{blue}{list\_ = len(list\_) + 0 \XSolidBrush}}\\ \\
    & \multirow{2}{*}{APT} &  & \multirow{2}{11.5cm}{\textcolor{red}{if} len(list\_) == 0: pass \textcolor{red}{\CheckmarkBold}}\\ \\
    \cdashline{2-4}
    & \multirow{2}{*}{TRANX} & \multirow{4}{2.5cm}{Close zfile stream.} & \multirow{2}{11.5cm}{\textcolor{blue}{close(zfile.stream()) \XSolidBrush}}\\ \\
    & \multirow{2}{*}{APT} &  & \multirow{2}{11.5cm}{\textcolor{red}{zfile}.close() \textcolor{red}{\CheckmarkBold}}\\ \\
    \midrule
    \multirow{8}{*}{CONALA} & \multirow{2}{*}{TRANX} & \multirow{4}{2.5cm}{Get attribute str\_0 from object var\_0.} & \multirow{2}{11.5cm}{\textcolor{blue}{print(var\_0.rstrip(str\_0)) \XSolidBrush}}\\ \\
    & \multirow{2}{*}{APT} &  & \multirow{2}{11.5cm}{\textcolor{red}{getattr}(var\_0, str\_0) \textcolor{red}{\CheckmarkBold}}\\ \\
    \cdashline{2-4}
    & \multirow{2}{*}{TRANX} & \multirow{4}{2.5cm}{Running bash script str\_0.} & \multirow{2}{11.5cm}{\textcolor{blue}{exec() \XSolidBrush}}\\ \\
    & \multirow{2}{*}{APT} &  & \multirow{2}{11.5cm}{\textcolor{red}{subprocess}.call(str\_0, shell=True) \textcolor{red}{\CheckmarkBold}}\\ \\
    \bottomrule
\end{tabular}}
\end{table*}

\subsection{AP Loss vs. Schedule Sampling} 
Schedule sampling is a conventional method employed to address the issue of inconsistency between the training and inference phases. In Table \ref{ss}, we compare AP Loss with schedule sampling, utilizing TRANX as the base model. It becomes evident that schedule sampling indeed results in improved performance on the training set during the inference phase, in terms of both BLEU and EM scores. However, schedule sampling does not significantly enhance the model's focus on antecedent predictions, leading to a relatively minor impact on the test set.

 \begin{figure}[h!]
\centering
\includegraphics[width=0.48\textwidth]{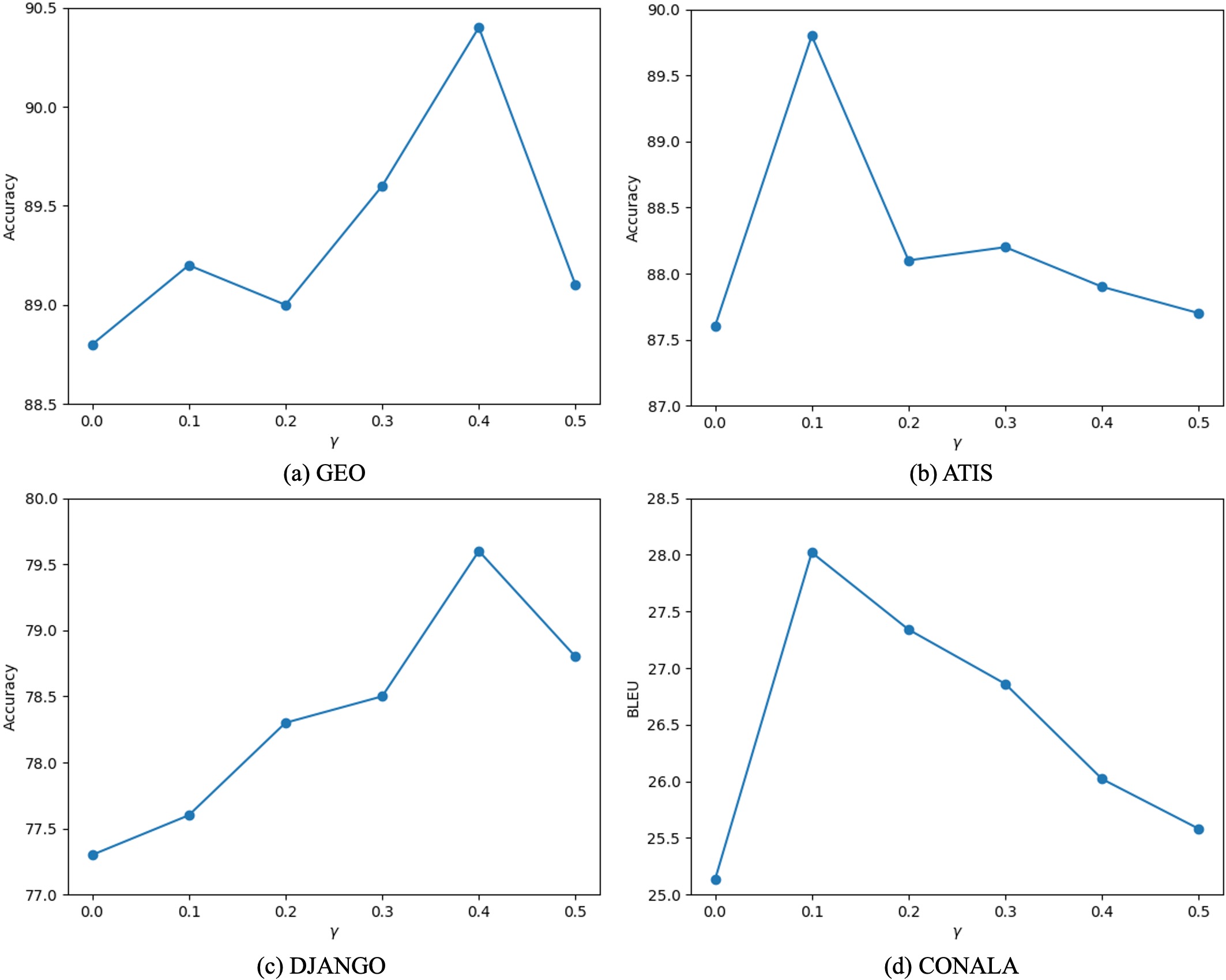}
\caption{
Performance of varying $\gamma$ in \eqref{pl} on four benchmark datasets.
}
\label{gamma}
\end{figure} 

\subsection{Effects of $\gamma$}
The coefficient $\gamma$ serves as a crucial hyperparameter for regulating AP Loss. To examine the effect of $\gamma$ on AP Loss, we present an analysis in Fig. \ref{gamma}. Optimal values for $\gamma$ are found to be 0.4 for GEO and DJANGO datasets, and 0.1 for ATIS and CONALA datasets. Interestingly, we observe an inverse correlation between the best $\gamma$ values and the average AST length, as presented in Table \ref{statistics}. This can be attributed to the fact that during the training phase, AP Loss must appropriately increase the scaling factor $f(\mathbf{a}, t)^{-\gamma}$ by reducing $\gamma$ for longer ASTs to prevent the neglect of later AST nodes.

\subsection{Case Study}
In Table \ref{table3}, we present a comparison of the top-1 generated code from TRANX and APT on various datasets. It is evident that the majority of errors tend to occur at the beginning or in the leading part of the code, subsequently leading to further faults. As illustrated in Table \ref{table3}, TRANX is prone to producing incorrect antecedent predictions, which poses a challenge for generating subsequent predictions while adhering to the constraints of AST. In contrast, our proposed model, APT, effectively mitigates the aforementioned issue by employing the AP Loss, thereby generating more accurate code.

\section{Related Work}
Recently, lots of code generation models have achieved excellent success \cite{Yin17,Alvarez-MelisJ17,hayati2018retrieval,SunZMXLZ19,Wei2019,xu2020incorporating,CODEP,CodeT5,CodeGPT,CodeScore,Self-planning,Self-collaboration}. The work \cite{Ling16} treated code generation as conditional text generation and solved it with the sequence-to-sequence model. In order to utilize the grammatical information of the code, the Seq2Tree models are resorted, transforming NL utterances into a sequence of AST-based grammar actions. Typically, Dong et al. \cite{DongL16} first explored a Seq2Tree model for code generation. The work \cite{RabinovichSK17} studied a modular encoder-decoder architecture with structured AST output spaces. Yin et al. \cite{TranX} proposed a Seq2Tree model to generate AST as the intermediate representation of codes. Moreover, the work \cite{TreeGen} used the Transformer \cite{attention17} architecture and attention mechanisms to address the long dependency problem for code generation. The authors in \cite{JiangSGMYS22} explored the use of the future nodes information generated by AST for prediction. The work \cite{IndustryCodeGeneration} proposed Subtoken-TranX adopted by Alibaba’s BizCook platform, which is the first domain code generation system adopted in industrial development environments. 

Some researchers have also noticed that the importance of each AST node for code generation is different. Xie et al. \cite{Xie21MLTranX} proposed a mutual learning framework for Seq2Tree models to learn the information of AST nodes in different traversals simultaneously. The work \cite{Jiang21TranXRL} selected the generation path of the node in AST using reinforcement learning. Significantly different from the above work, we explore the impact of the position information of AST nodes during the inference stage and propose an effective method to focus on the more critical antecedent nodes during the training stage. 

\section{Threats to Validity}
There are three major threats to the validity of our work. 

1) \textbf{Threats to external validity} concern the quality of experimental datasets and the generalizability of our results. First, the four public datasets for code generation are mainstream benchmarks and have been used in many related works \cite{TranX,Yin19Reranking,Xie21MLTranX,Jiang21TranXRL,JiangSGMYS22}. Second, AP Loss can be applied to any Seq2Tree model, and Seq2Seq models should further take into account the relationship between AST and code to employ AP Loss. In addition, APT uses only language-agnostic features and can be adapted to other programming languages.

2) \textbf{Threats to internal validity} involve the impact of hyperparameters. Deep learning models are known to be sensitive to hyperparameters. For the baselines, we work with the source code supplied by their original papers and ensure that the performance of the mode is comparable with their reported results. For our APT, we do a small-range grid search on hyperparameter $\gamma$, leaving other hyper-parameters the same as those in previous studies \cite{TranX,Yin19Reranking}. Previous work \cite{TranX,Yin19Reranking} has explored effective settings of the hyperparameters through extensive experiments. Therefore, there may be room to tune more hyperparameters for additional improvements. 

3) \textbf{Threats to construct validity} pertain to the reliability of our evaluation metrics. To address this threat, we employ EM and BLEU as evaluation metrics. EM evaluates the percentage of correctly predicted code snippets, and BLEU measures the text similarity between predictions and the ground truth. They are the mainstream metrics for code generation and are used in most previous studies \cite{TranX,Yin19Reranking,Xie21MLTranX,Jiang21TranXRL,JiangSGMYS22}. Based on the above metrics, each experiment is run five times, and its average result is reported.

\section{Conclusion and Discussion}
In this paper, we have identified the prediction of antecedent nodes as a critical factor influencing the behavior of Seq2Tree models. To address this issue, we have proposed an effective method, namely AP Loss, that adjusts the standard cross entropy loss to prioritize learning on antecedent nodes while down-weight the more susceptible subsequent nodes. Additionally, we have proposed an AST2Vec method that models AST using vectors to differentiate between antecedent and subsequent nodes. Leveraging AP Loss, APT outperforms existing Seq2Tree methods and achieves SOTA performance across four benchmarks. Comprehensive experimental results and analyses substantiate the effectiveness and versatility of our proposed methods. 

We demonstrate the efficacy of AP Loss by incorporating it into some widely-used Seq2Tree methods. However, the performance of APT is inherently constrained by the capabilities of the base model. In future work, we envisage applying AP Loss to training a larger Seq2Tree model following the method proposed by this paper, which could potentially lead to substantially superior outcomes.
Furthermore, we utilize AST2Vec to weigh AST nodes for AP Loss in this paper, and we intend to investigate the application of AST2Vec as a positional encoding in future research.

\section{Acknowledgments}
This research is supported by the National Natural Science Foundation of China under Grant Nos. 62192730, 62192733, 62192731, 61751210, 62072007, and 61832009.

\newpage
\appendix
\section{Supplementary Material}
\subsection{The Selection of $\alpha$}
\label{selection}
For continuous $f$, we can keep the magnitude of $\mathcal{L}_{\operatorname{ap}}$ equal to the magnitude of original $\mathcal{L}_{\operatorname{ce}}$ using the integral, which relates $\alpha$ to $\gamma$. For example, when $f(\mathbf{a}, t) = t$:
\begin{equation}
\left\{ 
\begin{aligned}
&\alpha = \frac{T}{\ln{(T+1)}}, \quad if\,\, \gamma = 1,\\
&\alpha = (1-\gamma) T^\gamma, \quad otherwise.
\end{aligned}
\right.
\label{a}
\end{equation}
With the help of \eqref{a}, we only need to tune $\gamma$, and the appropriate $\alpha$ will be calculated automatically. 

\begin{table}[h!]
	\caption{Varying $\alpha$ for AP Loss (Simple) with optimal $\gamma$, where `auto' denotes $\alpha$ is calculated by \eqref{a}.}	\label{table2}
	\centering
	\begin{tabular}{cccccc}
	    \toprule
        \multirow{2}{*}{$\alpha$} & GEO & ATIS & DJANGO & \multicolumn{2}{c}{CONALA}\\
        \cmidrule(r){2-2} \cmidrule(r){3-3}\cmidrule(r){4-4}\cmidrule(r){5-6}
         & EM & EM & EM & BLEU & EM\\ 
        \midrule
        1 & 89.6 & 87.9 & 78.8 & 26.75 & 2.4\\
        2 & \textbf{89.9} & 88.6 & 79.1 & \textbf{27.56} & 2.9\\
        4 & 88.8 & 87.7 & 78.5 & 25.64 & 2.6\\
        8 & 88.9 & 87.3 & 78.3 & 24.93 & 2.3 \\
        `auto' & 89.7 & \textbf{89.5} & \textbf{79.3} & 27.12 & \textbf{3.4}\\
	    \bottomrule
	\end{tabular}
  \end{table}

In Table \ref{table2}, we vary $\alpha$ for AP Loss (Simple) with optimal $\gamma$.  As $\alpha$ increases, the general trend of the performance of APT (Simple), i.e., the base model + AP Loss (Simple), rises and then falls on all datasets. When $\alpha$ is set to `auto', APT (Simple) achieves the best performance in terms of EM for ATIS, DJANGO, and CONALA, and impressive results regarding EM for GEO and BLEU for CONALA.
When we substitute the optimal $\gamma$ and the corresponding average AST length for each dataset into Eq. \eqref{a}, $\alpha$ is 1.96 for GEO, 1.27 for ATIS, 1.75 for DJANGO, and 1.23 for CONALA, which are all in the range of 1 to 2. Therefore, it is expected that the model works well when $\alpha$ is equal to 1 and 2. These results demonstrate the advantage of using Eq. \eqref{a} to calculate $\alpha$ automatically. 

\section{Computational Overhead}
Regarding the computational overhead of AP Loss, theoretically, the computation of AST2Vec is required with additional computational overhead during the training stage. However, in the data preparation stage, AST2Vec can be computed and stored while building the AST of the training data. This allows direct utilization of AST2Vec during the actual training stage, minimizing additional computational load. Furthermore, in the inference stage, our method incurs no additional computational overhead. Therefore, the additional computational overhead brought by the AP loss is very small. 

\newpage
\bibliography{sample-base}

\end{document}